%% file: skeleton.tex
\documentclass[a4paper,11pt]{article}
\usepackage{pos}
\usepackage{comment}
\usepackage{enumitem}
\usepackage{authorlist}
\usepackage{scalerel}
\usepackage{tikz}
\definecolor{orcidlogocol}{HTML}{A6CE39}

\title{KM3NeT/ARCA stacking search for high-energy neutrinos from blazars}

\author*[a,b]{Francesco Carenini}
\author[b]{Giulia Illuminati}
\onbehalf{on behalf of the KM3NeT Collaboration}
\affiliation[a]{Università di Bologna, Dipartimento di Fisica e Astronomia\\
Viale Berti Pichat 6/2, 40127 Bologna, Italia}

\affiliation[b]{INFN - Sezione di Bologna\\
Viale Berti Pichat 6/2, 40127 Bologna, Italia}

\emailAdd{francesco.carenini2@unibo.it}
\emailAdd{giulia.illuminati@bo.infn.it}

\abstract{
Blazars are promising targets for neutrino astronomy, as highlighted by IceCube’s identification of TXS 0506+056 as a cosmic neutrino source candidate. High-frequency-peaked BL Lacs (HBLs) stand out due to their characteristic electromagnetic high-energy emission properties, which makes them as well promising candidates for the production of high-energy neutrinos. Such neutrinos could be detected by KM3NeT/ARCA, a next-generation deep-sea Cherenkov detector under construction in the Mediterranean Sea. This contribution presents the results of a binned likelihood stacking analysis investigating high-energy neutrino emissions from a subset of HBLs. A total of 232 sources have been selected from the 3HSP catalogue and targeted in the analysis of KM3NeT/ARCA data taken with several detector configurations, specifically in the period in which from 6 to 21 detection units were deployed. The analysis is based on neutrino emission models of blazars developed using the LeHa-Paris numerical code, which simulates the spectral energy distribution of jetted active galactic nuclei.}

\ConferenceLogo{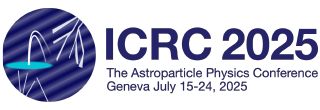}

\FullConference{39th International Cosmic Ray Conference (ICRC2025)\\
 15–24 July 2025\\
Geneva, Switzerland\\}
\begin{document}
\maketitle

\section{Introduction}
High-energy neutrinos offer a unique probe of hadronic processes in the most extreme astrophysical environments. Their production is expected from the decay of charged mesons originating in proton-proton or proton-photon interactions at cosmic accelerators. One prominent class of candidate sources are blazars, i.e. active galactic nuclei with relativistic jets pointing towards Earth. IceCube has reported the detection of a high-energy neutrino coincident with a gamma-ray flare from the blazar TXS 0506+056 \cite{IceCube:2018dnn}, and further evidence for past neutrino activity from the same source \cite{IceCube:2018cha}, suggesting that blazars may contribute to the observed diffuse astrophysical neutrino flux. Next-generation deep-sea neutrino telescopes, such as KM3NeT/ARCA, are expected to advance the field significantly. KM3NeT/ARCA \cite{KM3Net:2016zxf} is a Cherenkov neutrino telescope currently under construction in the Mediterranean Sea, designed to detect high-energy neutrinos with good angular resolution in a wide energy range. It consists of an array of multi-PMT optical modules~\cite{Ref4} arranged along vertical strings (Detection Units, or DUs) anchored to the sea floor, optimized for the detection of Cherenkov light induced by relativistic charged particles produced in neutrino interactions in water. It will feature a cubic-kilometer detection volume and target high-energy astrophysical neutrinos, with energies from 100 GeV up to multi-PeV. Thanks to its modular design, data taking is already possible in the construction phase. This work introduces a binned likelihood stacking analysis designed to search for high-energy neutrino emission from high-frequency peaked BL Lac objects (HBLs) using data taken with KM3NeT/ARCA, in configurations with 6, 8, 19, and 21 DUs. Neutrino flux spectra have been derived in a one-zone lepto-hadronic scenario, using the LeHa-Paris code \cite{Cerruti:2014iwa}, that simulates radiative processes in jets of supermassive-black-holes and the associated photo-meson interactions. Specifically, this contribution looks at a selected subset of HBLs from the 3HSP catalogue \cite{Chang:2019vfd}.

\section{Neutrinos from blazars}
\label{sec:2}
In one-zone lepto-hadronic models, a magnetized compact region within the relativistic jet contains co-accelerated populations of relativistic electrons and protons. Neutrinos are produced through the decay of pions generated in proton-photon interactions. The high-energy component of the spectral energy distribution (SED) can arise, in addition to synchrotron-self-Compton from primary electrons (the leptonic contribution), also from synchrotron-pair cascades of secondary particles, neutral pion decay or proton synchrotron radiation (the hadronic contribution). The low-energy SED component is attributed to synchrotron emission from primary electrons. In this study, as a first step, the neutrino spectrum of the blazar PKS 2155-304 has been computed using LeHa-Paris. PKS 2155-304 is a HBL object located in the Southern Hemisphere ($\delta = -30.22^{\circ}$) at redshift $z = 0.117$. Selected as a representative HBL candidate, its SED has been studied by exploring a broad parameter space and comparing the results with multi-wavelength observational data from \cite{Madejski:2016evb}. The optimization of the modeling is detailed in \cite{careniniproc}, with the choice of the following parameters yielding the best-fit neutrino flux: an electron normalization $K_e$, at Lorentz factor \( \gamma = 1 \), of \(1.3 \times 10^4 \ \text{cm}^{-3} \); a maximum proton energy of \( \log_{10}(\gamma_p^{\text{max}}) = 7.4 \); a proton-to-electron ratio at \( \gamma = 1 \) of \( \eta = 0.003 \); a spectral index for proton acceleration \( (\alpha_p) \) corresponding to 1.8; a magnetic field of \( B = 0.036 \ \text{G} \); an emission region of size \( R \sim 7.9 \times 10^{16} \ \text{cm} \) and a break in the electron stationary energy distribution at \( \gamma_{\text{break}} = 6.3 \times 10^4 \).  Additionally, a Doppler factor \( \delta = 33 \) and a viewing angle of \( 0.1^\circ \) (corresponding to a bulk Lorentz factor \( \Gamma \sim 16 \)) have been assumed.

\subsection{Catalogue definition}
\begin{figure}[h!]
    \centering
    \includegraphics[width=0.64\linewidth]{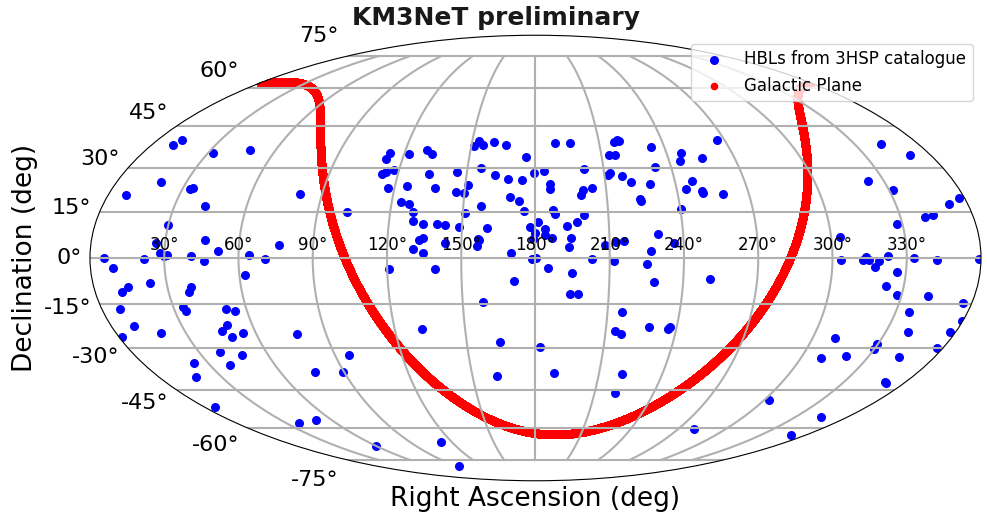}
    \caption{Sky map in equatorial coordinates showing the selected HBLs sources from the 3HSP catalogue. A total of 232 sources satisfy the selection criteria. The Galactic Plane is indicated in red.}
    \label{fig:skymap}
\end{figure}

Once the best-fit flux for PKS 2155-304 was identified, the modeling has been extended to a larger population of selected HBLs blazars from the 3HSP catalogue. The 3HSP catalogue \cite{Chang:2019vfd} provides the most extensive list of High Synchrotron Peaked and Extreme HBLs to date, for a total of 2013 candidates. For the purposes of this analysis, sources were selected according to the following criteria: requiring a secure source identification, a firm redshift measurement, a declination less than 40 degrees, to account for the visibility of KM3NeT/ARCA, and a synchrotron peak frequency in the range $15.0 < \log(\nu^{\text{syn}}_{\text{peak}}/\text{Hz}) < 16.8$. This frequency range was adopted in line with the standard definition of HBLs. These selection criteria resulted in 232 selected HBLs, whose distribution in the sky is shown in Figure \ref{fig:skymap}. 

\subsection{Neutrino fluxes}
\begin{figure}[h!]
    \centering
    \includegraphics[width=0.74\linewidth]{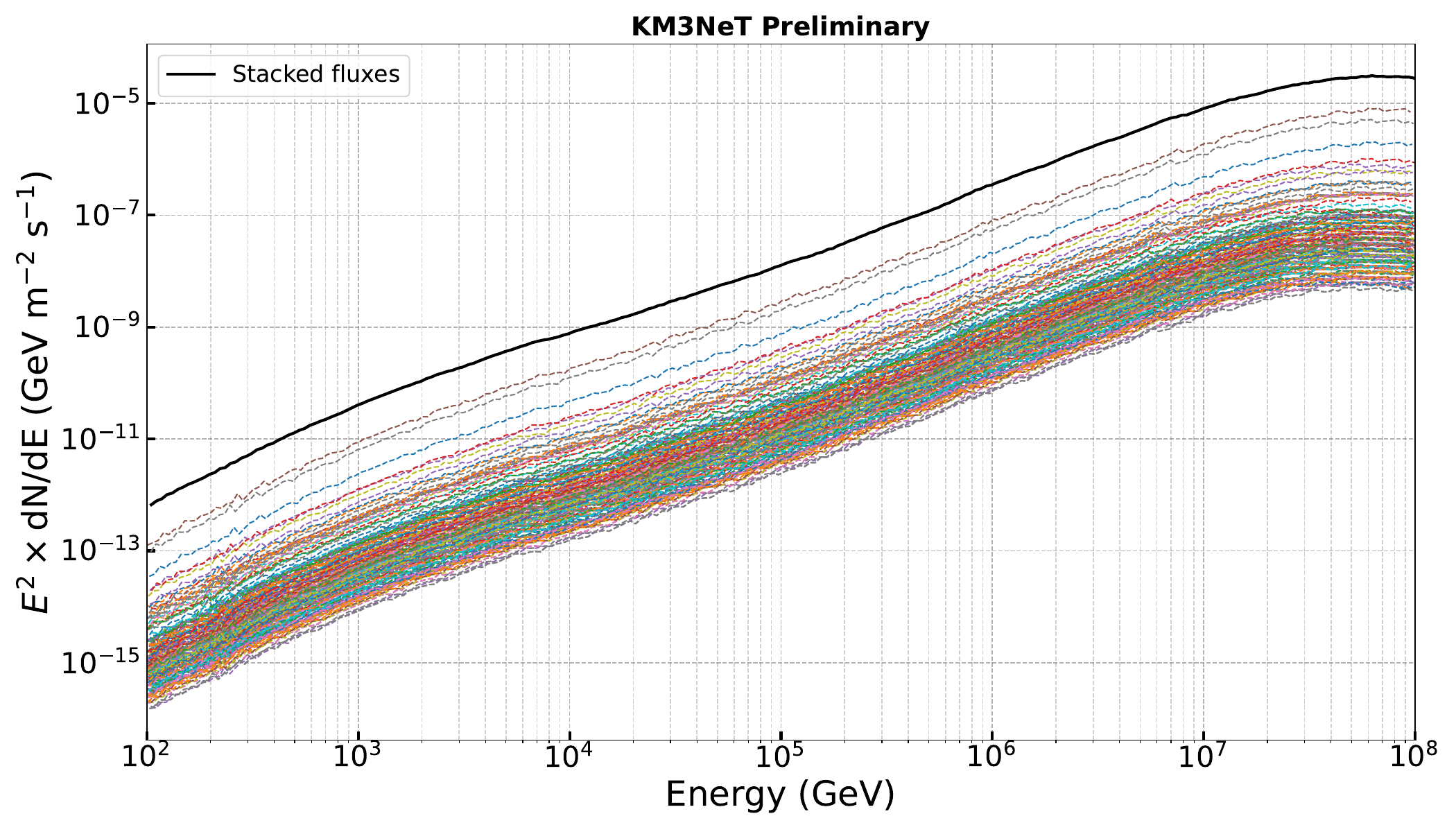}
    \caption{Expected neutrino fluxes for the 232 selected HBLs (dashed colored lines) and their stacked flux (solid black line). Both neutrino and antineutrino separation and their oscillated flavor composition at Earth are considered in the calculation.}
    \label{fig:template}
\end{figure}

Rather than re-optimizing the model parameters for each individual source, the expected neutrino flux for each blazar is obtained by rescaling the neutrino best-fit flux of PKS 2155-304 according to source-specific observables reported in the catalogue, thereby making the model predictive for a larger sample. Specifically, the synchrotron peak luminosity serves as a proxy for the neutrino luminosity. This is motivated by the fact that synchrotron emission arises from the electron population, which is co-accelerated with protons under the same physical conditions. A more luminous synchrotron peak implies a denser photon field, and consequently, a potentially higher neutrino production efficiency. The neutrino energy, instead, scales according to the redshift of the source under consideration. Thus, following the methods described in \cite{careniniproc}, neutrino fluxes for the 232 selected HBLs have been derived: their stacked contribution is depicted in Figure \ref{fig:template} together with the predicted fluxes of individual blazars. The fluxes are displayed in the energy range from 100 GeV to 100 PeV, relevant for KM3NeT/ARCA. The calculation accounts for neutrinos and antineutrinos, as well as their mixed flavor composition at Earth due to neutrino oscillations over cosmic distances. Because of the adopted rescaling procedure, all the templates share a common spectral shape, featuring a pronounced peak at the level of $10^{17}$ eV.

\section{Data sample and event selection}

This analysis focuses on data collected with the KM3NeT/ARCA detector operating in various configurations from 6 to 21 instrumented strings. The dataset comprises approximately 92.0 days with 6 DUs (ARCA6), 212.3 days with 8 DUs (ARCA8), 48.4 days with 19 DUs (ARCA19), and about 287.4 days with 21 DUs (ARCA21), totaling a livetime of roughly 640 days. Each dataset is associated with corresponding instrument response functions (IRFs), which are essential for signal simulations, while backgrounds are derived directly from the data itself, as described in the following. The event selection criteria follow those adopted for the point-source search developed with KM3NeT/ARCA6–21 \cite{parisi}, considering track-like events, i.e. the results of mainly charged-current muon neutrino interactions. Cuts are applied to suppress atmospheric muon contamination and to exclude poorly reconstructed events. These selections employ cuts on the number of hits (a hit is defined as a PMT anode signal above a certain threshold, typically 0.3 photo-electrons) used in the reconstruction, the reconstructed track direction and the reconstruction quality, which is quantified by a likelihood-based fit parameter. For the ARCA19–21 period, the event selection was further improved by applying cuts on the reconstructed track length and the angular uncertainty of the reconstructed direction, as well as by training a boosted decision tree (BDT) classifier to enhance the separation between signal and background events. Overall, 17285 events are selected.

\section{Background and signal modeling}
\label{sec:4}
The background rate as a function of reconstructed energy and declination ($\delta$) is estimated from scrambled data, i.e. data randomized in right ascension. A factorized model is employed to describe the background’s dependence on energy and declination. The expected background event rate per unit solid angle and logarithm of the energy is given by:

\begin{equation}
N_{\text{bkg}} = n \cdot F(\delta) \cdot F(\log_{10}(E)) \quad \text{[sr}^{-1} \cdot (\log_{10}(\text{GeV}))^{-1}\text{]},
\end{equation}

where the normalization constant $n$ is chosen such that the integral of $N_{\text{bkg}}$ over the full sky and energy range matches the total number of observed events in each dataset. The declination dependence $F(\delta)$ is modeled using a spline fit in $\sin(\delta)$, while the energy dependence $F(\log_{10}(E))$ is parameterized using a sum of Gaussian functions. This estimation approach is validated by previous ANTARES and KM3NeT/ARCA analyses \cite{Dornic:2023Dt}.\\
The detector response to a potential neutrino signal is modeled using Monte Carlo simulations and comprises three main components: the effective area, that depends on the true neutrino energy and its arrival direction, and quantifies the equivalent area over which the detector is fully efficient at detecting neutrinos of a given energy and direction; the energy resolution, given by the distribution of reconstructed energies for a given true energy and direction; the point spread function (PSF), that gives the probability of reconstructing a neutrino at a certain distance from the source location. These IRFs are used to generate background and signal probability density functions in the form of 2D-histograms, containing the logarithm of the reconstructed energy of the event and the distance of the reconstructed track to the candidate source, as described below.

\section{Stacking analysis}

\subsection{Likelihood formalism}

The core of the analysis relies on a binned maximum likelihood formalism. Reconstructed tracks are binned in two dimensions: the angular distance $\alpha$ to the source, in the range [0;5] degrees, and the reconstructed energy $\log_{10}(E_{\mathrm{rec}})$, in the range [1;8] (in $\log_{10}(\mathrm{Energy/GeV})$). For each bin $i$, the expected number of background events $B_i$ and the expected signal $S_i$ (for the fluxes described in Section \ref{sec:2}) are defined according to the ingredients outlined in Section \ref{sec:4}.

The log-likelihood function for a given dataset is expressed as:
\begin{equation}
\log \mathcal{L} = \sum_{i \in \text{bins}} N_i \log(B_i + \zeta S_i) - B_i - \zeta S_i,
\end{equation}
where $\zeta$ is the signal strength, a scaling parameter for the reference flux, and $N_i$ is the observed number of events. The optimal signal strength $\hat{\zeta}$ is determined by maximizing $\log \mathcal{L}$.

The test statistic (TS) is defined as the logarithm of the likelihood ratio:
\begin{equation}
\text{TS} = \log \left( \frac{\mathcal{L}(\zeta = \hat{\zeta})}{\mathcal{L}(\zeta = 0)} \right),
\end{equation}

and is used to quantify the preference for the signal hypothesis (H1) over the background-only one (H0). To derive statistical significance and upper limits, pseudo-experiments (PEs) are generated by randomly drawing $N_i$ from a Poisson distribution with its mean given by $B_i + \zeta_{\text{true}} S_i$, where $\zeta_{\text{true}}$ is the true signal strength.

\subsection{Stacking Analysis Methodology}

\begin{figure}[h!]
\centering
\includegraphics[width=0.68\textwidth]{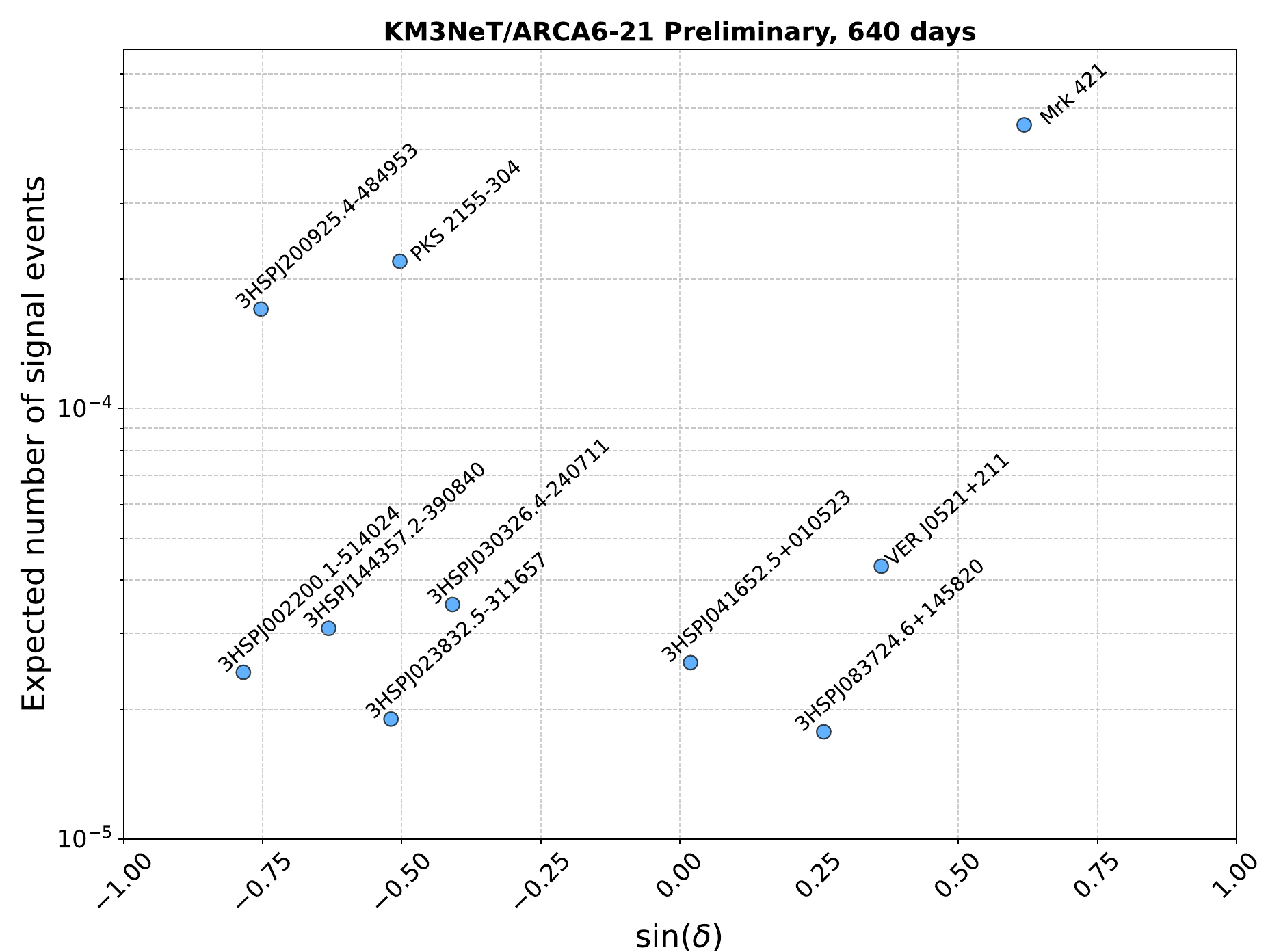}
\caption{Top 10 brightest sources, among the selected HBLs of the 3HSP catalogue, ranked by the expected neutrino yield in the KM3NeT/ARCA detector. The expected number of signal events in KM3NeT/ARCA6-21 is shown as a function of the sine of the source declination ($\delta$).}
\label{fig:top10_sources}
\end{figure}

To increase sensitivity to weak signals across multiple sources, a stacking analysis is implemented. The 232 source candidates are ranked by their respectively expected neutrino yield given by the assumed spectral models (see Section \ref{sec:2}). Among the 10 sources that contribute the most to the expected signal (Figure~\ref{fig:top10_sources}), are the nearby HBL Mrk 421 \cite{Balokovic:2016}, which is an outlier of this population, PKS 2155-304 and the gamma-ray emitter VER J0521+211 \cite{VERITAS_MAGIC:2022}. Following this ordering, a point-source analysis is carried out individually for each candidate source to obtain their respective likelihood profiles. These profiles are subsequently combined to
perform a joint fit. This is achieved summing the individual likelihoods across the catalogue to construct a global likelihood function:
\begin{equation}
\log \mathcal{L}_{\text{combined}}(\zeta) = \sum_i \log \mathcal{L}_i(\zeta),
\end{equation}
where $i$ indexes the sources. This approach preserves the relative contribution of each source and allows for a single signal strength parameter $\zeta$ to be fitted globally.

PEs are carried out separately for the H0 and H1 hypothesis.  Specifically, $10^5$ PEs were generated for H0, while $2 \cdot 10^4$ PEs were simulated for H1. From the resulting distribution of TS values, discovery potentials (DPs) and sensitivities are then computed using the same technique as implemented in KM3NeT/ARCA's binned point-source analysis \cite{parisi}. To obtain a reliable estimate of DPs, the tail of the TS distribution under the H0 hypothesis is fitted with an exponential function. This approach enables a more accurate integration than direct histogram-based methods.

\subsection{Flux uncertainty and systematics}
\label{sub:fluxunc}
A conservative band spanning one order of magnitude around the flux, based on the PKS~2155-304 rescaling, is adopted for each source. This choice accounts for intrinsic limitations of the rescaling method, which assumes identical proton content and similar physical conditions across all sources in the catalogue. In fact, the model validation (see \cite{careniniproc}), using Mrk 421 and VER~J0521+211, two of the dominant contributors to the stacking, shows agreement within a factor of 5 and 10, respectively, in the energy range where 90\% of the signal is expected in KM3NeT/ARCA6-21 for the stacked fluxes, approximately from 6.1 to 7.8 in $\log_{10}(\mathrm{Energy/GeV})$. 
The factor-10 band is therefore an estimate that also encompasses other catalogue sources. 
For instance, the source ranking fifth in terms of flux, 3HSPJ030326.4-240711 (see Figure~\ref{fig:template}), is approximately a factor of 10 fainter than Mrk 421, in the aforementioned energy range. Hence, even if the flux was misestimated by about a factor of 10, it would still lie within the adopted uncertainty band and remain subdominant in the overall stacked signal. The stacked neutrino flux and its uncertainty are shown in Figure~\ref{fig:UL}. Limits are computed assuming the central flux in the likelihood, while the uncertainty band illustrates the range of possible stacked flux values.\\
Regarding systematic uncertainties on detector performance, they are included by applying a $0.5^\circ$ Gaussian smearing to the PSF and a 30\% Gaussian spread to the acceptance, following the implementation in point-source analyses \cite{parisi}.

\section{Results}

\begin{figure}[h!]
\centering
\includegraphics[width=0.78\textwidth]{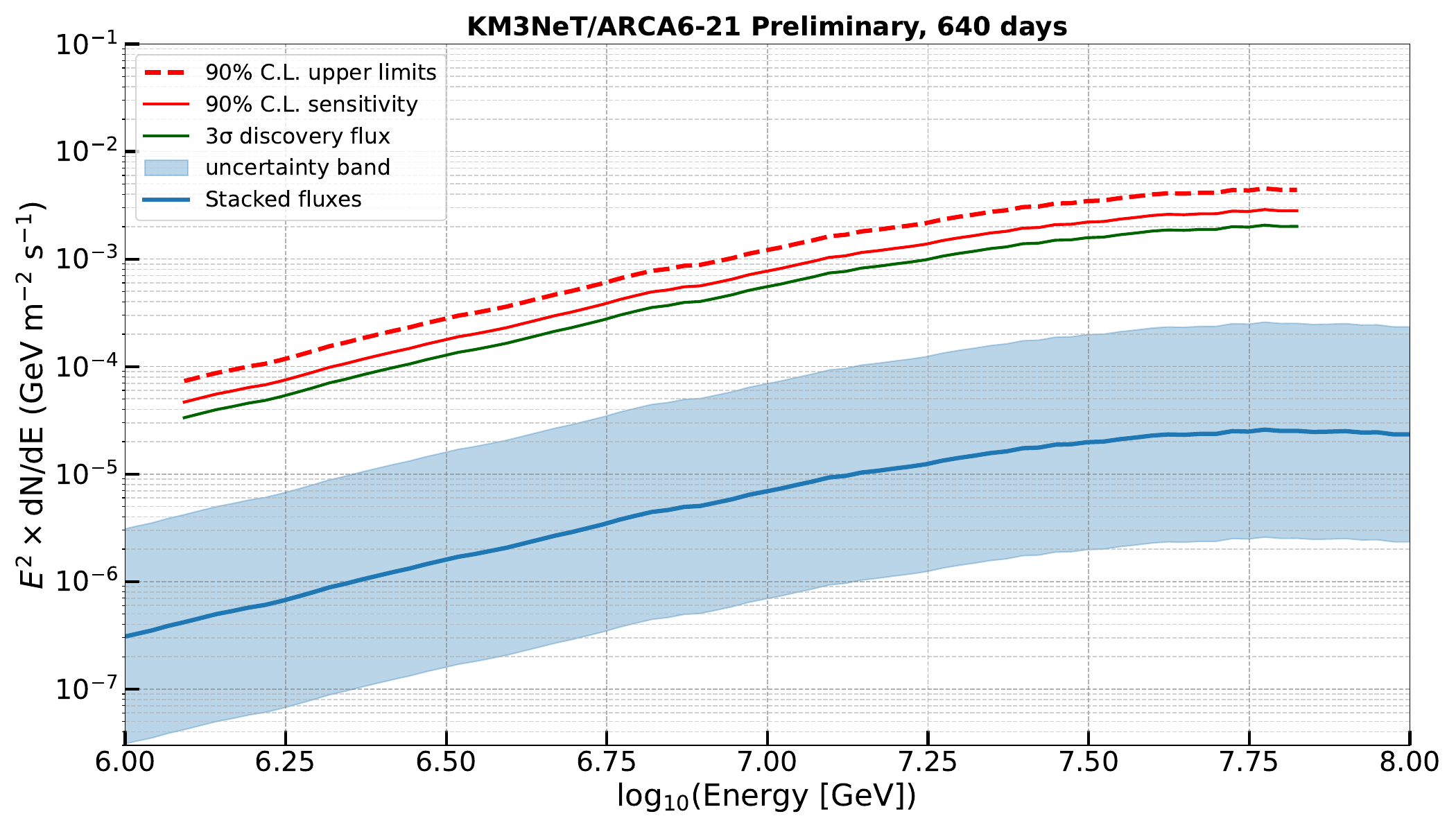}
\caption{Sensitivity at 90\% confidence level (solid red line) and discovery flux at $3\sigma$ significance (solid green line) for the stacked flux (solid blue), shown in the energy range where 90\% of the signal is expected.  The shaded light-blue region indicates the flux uncertainty. The upper limit on the central flux is indicated as a red dashed line.}
\label{fig:UL}
\end{figure}

The analysis was extended by progressively adding sources from the catalogue, to investigate the improvement in sensitivity with an increasing number of stacked sources. After the inclusion of the first 10 sources, the incremental improvement from adding new sources became less significant, reaching a plateau at approximately 100 sources, beyond which further additions yielded negligible gains. Those 100 sources have been unblinded. At this saturation point, the sensitivity and discovery flux for the KM3NeT/ARCA6-21 full dataset were computed within the energy range encompassing 90\% of the expected signal. Results are shown in Figure~\ref{fig:UL}.

Both the 90$\%$ confidence level sensitivity and the $3\sigma$ discovery flux lie approximately two orders of magnitude above the stacked central flux. The associated flux uncertainty described in \ref{sub:fluxunc} is also shown.\\
After unblinding, no significant excess of events was observed over the background expectation. The calculated p-value is 0.074, corresponding to a statistical significance of 1.44$\sigma$, in a one-sided convention. Then, 90\% confidence level upper limits on the stacked flux have been derived and are shown in Figure \ref{fig:UL}.
\section{Conclusions}
This work presented a stacking analysis targeting high-frequency-peaked BL Lac objects selected from the 3HSP catalogue. It relied on neutrino emission models computed with the LeHa-Paris code and exploited data collected by KM3NeT/ARCA, with 6, 8, 19 and 21 DUs. Stacking analyses are known to enhance the sensitivity to faint neutrino signals from astrophysical source populations. Although no significant excess has been observed, the use of physically motivated models strengthens the connection between theoretical predictions and data. With the ongoing expansion of the deep-sea infrastructure of KM3NeT/ARCA, future analyses with increased exposure and refined reconstruction capabilities are expected to improve the results.

\begingroup
\fontsize{11}{10}\selectfont

\endgroup

\newpage
\input{KM3NeT_PUB_ICRC25_v1_modified}

\end{document}

%% file: KM3NeT_PUB_ICRC25_v1_modified.tex
\section*{Full Authors List: The KM3NeT Collaboration}

\scriptsize
\noindent
{O.~Adriani\,\orcidlink{0000-0002-3592-0654}}$^{b,a}$,
{A.~Albert}$^{c,be}$,
{A.\,R.~Alhebsi\,\orcidlink{0009-0002-7320-7638}}$^{d}$,
{S.~Alshalloudi}$^{d}$,
{M.~Alshamsi}$^{e}$,
{S. Alves Garre\,\orcidlink{0000-0003-1893-0858}}$^{f}$,
{F.~Ameli}$^{g}$,
{M.~Andre}$^{h}$,
{L.~Aphecetche\,\orcidlink{0000-0001-7662-3878}}$^{i}$,
{M. Ardid\,\orcidlink{0000-0002-3199-594X}}$^{j}$
% ...
\\ \\
$^{a}${Universit{\`a} di Firenze, Dipartimento di Fisica e Astronomia, via Sansone 1, Sesto Fiorentino, 50019 Italy}\\
$^{b}${INFN, Sezione di Firenze, via Sansone 1, Sesto Fiorentino, 50019 Italy}\\
$^{c}${Universit{\'e}~de~Strasbourg,~CNRS,~IPHC~UMR~7178,~F-67000~Strasbourg,~France}\\
$^{d}${Khalifa University of Science and Technology, Department of Physics, PO Box 127788, Abu Dhabi,   United Arab Emirates}\\
$^{e}${Aix~Marseille~Univ,~CNRS/IN2P3,~CPPM,~Marseille,~France}\\
$^{f}${IFIC - Instituto de F{\'\i}sica Corpuscular (CSIC - Universitat de Val{\`e}ncia), c/Catedr{\'a}tico Jos{\'e} Beltr{\'a}n, 2, 46980 Paterna, Valencia, Spain}\\
$^{g}${INFN, Sezione di Roma, Piazzale Aldo Moro, 2 - c/o Dipartimento di Fisica, Edificio, G.Marconi, Roma, 00185 Italy}\\
$^{h}${Universitat Polit{\`e}cnica de Catalunya, Laboratori d'Aplicacions Bioac{\'u}stiques, Centre Tecnol{\`o}gic de Vilanova i la Geltr{\'u}, Avda. Rambla Exposici{\'o}, s/n, Vilanova i la Geltr{\'u}, 08800 Spain}\\
$^{i}${Subatech, IMT Atlantique, IN2P3-CNRS, Nantes Universit{\'e}, 4 rue Alfred Kastler - La Chantrerie, Nantes, BP 20722 44307 France}\\
$^{j}${Universitat Polit{\`e}cnica de Val{\`e}ncia, Instituto de Investigaci{\'o}n para la Gesti{\'o}n Integrada de las Zonas Costeras, C/ Paranimf, 1, Gandia, 46730 Spain}\\
% ...
$^{be}${Universit{\'e} de Haute Alsace, rue des Fr{\`e}res Lumi{\`e}re, 68093 Mulhouse Cedex, France}\\
% ...

\section*{Acknowledgements}

\noindent The authors acknowledge the financial support of:
KM3NeT-INFRADEV2 project, funded by the European Union Horizon Europe Research and Innovation Programme under grant agreement No 101079679;
Funds for Scientific Research (FRS-FNRS), Francqui foundation, BAEF foundation.
Czech Science Foundation (GAČR 24-12702S);
Agence Nationale de la Recherche (contract ANR-15-CE31-0020), Centre National de la Recherche Scientifique (CNRS), Commission Europ\'eenne (FEDER fund and Marie Curie Program), LabEx UnivEarthS (ANR-10-LABX-0023 and ANR-18-IDEX-0001), Paris \^Ile-de-France Region, Normandy Region (Alpha, Blue-waves and Neptune), France,
The Provence-Alpes-Côte d'Azur Delegation for Research and Innovation (DRARI), the Provence-Alpes-Côte d'Azur region, the Bouches-du-Rhône Departmental Council, the Metropolis of Aix-Marseille Provence and the City of Marseille through the CPER 2021-2027 NEUMED project,
The CNRS Institut National de Physique Nucléaire et de Physique des Particules (IN2P3);
Shota Rustaveli National Science Foundation of Georgia (SRNSFG, FR-22-13708), Georgia;
This research was funded by the European Union (ERC MuSES project No 101142396);
The General Secretariat of Research and Innovation (GSRI), Greece;
Istituto Nazionale di Fisica Nucleare (INFN) and Ministero dell’Universit{\`a} e della Ricerca (MUR), through PRIN 2022 program (Grant PANTHEON 2022E2J4RK, Next Generation EU) and PON R\&I program (Avviso n. 424 del 28 febbraio 2018, Progetto PACK-PIR01 00021), Italy; IDMAR project Po-Fesr Sicilian Region az. 1.5.1; A. De Benedittis, W. Idrissi Ibnsalih, M. Bendahman, A. Nayerhoda, G. Papalashvili, I. C. Rea, A. Simonelli have been supported by the Italian Ministero dell'Universit{\`a} e della Ricerca (MUR), Progetto CIR01 00021 (Avviso n. 2595 del 24 dicembre 2019); KM3NeT4RR MUR Project National Recovery and Resilience Plan (NRRP), Mission 4 Component 2 Investment 3.1, Funded by the European Union – NextGenerationEU,CUP I57G21000040001, Concession Decree MUR No. n. Prot. 123 del 21/06/2022;
Ministry of Higher Education, Scientific Research and Innovation, Morocco, and the Arab Fund for Economic and Social Development, Kuwait;
Nederlandse organisatie voor Wetenschappelijk Onderzoek (NWO), the Netherlands;
The grant “AstroCeNT: Particle Astrophysics Science and Technology Centre”, carried out within the International Research Agendas programme of the Foundation for Polish Science financed by the European Union under the European Regional Development Fund; The program: “Excellence initiative-research university” for the AGH University in Krakow; The ARTIQ project: UMO-2021/01/2/ST6/00004 and ARTIQ/0004/2021;
Ministry of Education and Scientific Research, Romania;
Slovak Research and Development Agency under Contract No. APVV-22-0413; Ministry of Education, Research, Development and Youth of the Slovak Republic;
MCIN for PID2021-124591NB-C41, -C42, -C43 and PDC2023-145913-I00 funded by MCIN/AEI/10.13039/501100011033 and by “ERDF A way of making Europe”, for ASFAE/2022/014 and ASFAE/2022 /023 with funding from the EU NextGenerationEU (PRTR-C17.I01) and Generalitat Valenciana, for Grant AST22\_6.2 with funding from Consejer\'{\i}a de Universidad, Investigaci\'on e Innovaci\'on and Gobierno de Espa\~na and European Union - NextGenerationEU, for CSIC-INFRA23013 and for CNS2023-144099, Generalitat Valenciana for CIDEGENT/2020/049, CIDEGENT/2021/23, CIDEIG/2023/20, ESGENT2024/24, CIPROM/2023/51, GRISOLIAP/2021/192 and INNVA1/2024/110 (IVACE+i), Spain;
Khalifa University internal grants (ESIG-2023-008, RIG-2023-070 and RIG-2024-047), United Arab Emirates;
The European Union's Horizon 2020 Research and Innovation Programme (ChETEC-INFRA - Project no. 101008324).

Views and opinions expressed are those of the author(s) only and do not necessarily reflect those of the European Union or the European Research Council. Neither the European Union nor the granting authority can be held responsible for them.